\documentclass[aps,prl,twocolumn,superscriptaddress]{revtex4-1}
\usepackage{graphicx}
\usepackage{lmodern}

\usepackage{dcolumn}
\usepackage{bm}
\usepackage{color}
\usepackage{ulem}
\usepackage[usenames,dvipsnames,svgnames,table]{xcolor}

\newcommand{\pstar}{$p^{\star}$}
\newcommand{\Tstar}{$T^{\star}$}

\newcommand{\mstar}{$m^{\star}$}

\newcommand{\nH}{$n_{\rm H}$}

\newcommand{\Tc}{$T_{\rm c}$}

\newcommand{\Hc}{$H_{\rm c2}$}

\newcommand{\Hvs}{$H_{\rm vs}$}

\begin{document}



\title{High density of states in the pseudogap phase of the cuprate superconductor HgBa$_2$CuO$_{4 + \delta}$
from low-temperature normal-state specific heat}

\author{C.~Girod}
\affiliation{Univ. Grenoble Alpes, CNRS, Grenoble INP, Institut N\'eel, F-38000 Grenoble, France}
\affiliation{Institut Quantique, D\'epartement de physique \& RQMP, Universit\'e de Sherbrooke, Sherbrooke, Qu\'ebec J1K 2R1, Canada}

\author{A.~Legros}
\affiliation{Institut Quantique, D\'epartement de physique \& RQMP, Universit\'e de Sherbrooke, Sherbrooke, Qu\'ebec J1K 2R1, Canada}
\affiliation{SPEC, CEA, CNRS-UMR3680, Universit\'e Paris-Saclay, Gif-sur-Yvette Cedex 91191, France}

\author{A.~Forget}
\affiliation{SPEC, CEA, CNRS-UMR3680, Universit\'e Paris-Saclay, Gif-sur-Yvette Cedex 91191, France}

\author{D.~Colson}
\affiliation{SPEC, CEA, CNRS-UMR3680, Universit\'e Paris-Saclay, Gif-sur-Yvette Cedex 91191, France}

\author{C.~Marcenat}
\affiliation{Univ. Grenoble Alpes, CEA, IRIG, PHELIQS, LATEQS, F-38000 Grenoble, France}

\author{A.~Demuer}
\affiliation{Univ. Grenoble Alpes, INSA Toulouse, Univ. Toulouse Paul Sabatier, CNRS, LNCMI, F-38000 Grenoble, France}

\author{D.~LeBoeuf}
\affiliation{Univ. Grenoble Alpes, INSA Toulouse, Univ. Toulouse Paul Sabatier, CNRS, LNCMI, F-38000 Grenoble, France}

\author{L.~Taillefer}
\email[]{louis.taillefer@usherbrooke.ca}
\affiliation{Institut Quantique, D\'epartement de physique \& RQMP, Universit\'e de Sherbrooke, Sherbrooke, Qu\'ebec J1K 2R1, Canada}
\affiliation{Canadian Institute for Advanced Research, Toronto, Ontario  M5G 1Z8, Canada}

\author{T.~Klein}
\email[]{thierry.klein@neel.cnrs.fr}
\affiliation{Univ. Grenoble Alpes, CNRS, Grenoble INP, Institut N\'eel, F-38000 Grenoble, France}

\date{\today}

\begin{abstract}

The specific heat $C$ of the single-layer cuprate superconductor HgBa$_2$CuO$_{4 + \delta}$ was measured in an underdoped crystal 
with \Tc~$= 72$~K at temperatures down to 2~K in magnetic fields up to 35~T, a field large enough to suppress superconductivity 
at that doping ($p \simeq 0.09$).
In the normal state at $H = 35$~T, a residual 
linear term of magnitude $\gamma = 12 \pm 2$~mJ/K$^2$mol is observed in $C/T$ as $T \to 0$,
a direct measure of the electronic density of states.
This high value of $\gamma$ has two major implications.
First, it is significantly larger than the value measured in overdoped cuprates outside the pseudogap phase
($p >$~\pstar),
such as 
La$_{2-x}$Sr$_x$CuO$_4$
and
Tl$_2$Ba$_2$CuO$_{6 + \delta}$
at $p \simeq 0.3$, where $\gamma \simeq 7$~mJ/K$^2$mol.
Given that the pseudogap causes a loss of density of states, 
and assuming that HgBa$_2$CuO$_{4 + \delta}$ has the same $\gamma$ value as other cuprates at $p \simeq 0.3$,
this implies that $\gamma$ 
in HgBa$_2$CuO$_{4 + \delta}$
must  
peak between $p \simeq 0.09$ and $p \simeq 0.3$, namely at (or near) the  critical doping \pstar~where
the pseudogap phase is expected to end (\pstar~$\simeq 0.2$).
Secondly, the high $\gamma$ value implies that the Fermi surface must consist of more than the single electron-like 
pocket detected by quantum oscillations in HgBa$_2$CuO$_{4 + \delta}$ at $p \simeq 0.09$,
whose effective mass 
\mstar~$= 2.7~m_0$~yields only $\gamma = 4.0$~mJ/K$^2$mol.
This missing mass imposes a revision of the current scenario for how pseudogap and charge order 
respectively transform and reconstruct the Fermi surface of cuprates.

\end{abstract}

\pacs{}

\maketitle


\section{I.~Introduction}

Despite three decades of intense research, fundamental questions remain about the 
phase diagram of cuprate superconductors~\cite{Keimer2015}.
The central enigma is the nature of the pseudogap phase,
an elusive phase that exists below a temperature \Tstar~and below a critical hole concentration (doping)
\pstar~(Fig.~1), whose defining characteristic is a drop in the electronic density of states (DOS)~\cite{Proust2019}.
To crack this enigma, a crucial piece of information is the Fermi surface in the ground state of the pseudogap phase,
at $T=0$ without superconductivity, and the associated DOS.
This kind of information has only recently begun to surface~\cite{Michon2019}, but the picture is still far from complete.

Well above \pstar, cuprates are fairly conventional metals with a well characterized Fermi surface, namely a large quasi-2D cylinder, 
in agreement with band structure calculations.
In the single-layer material Tl$_2$Ba$_2$CuO$_{6+\delta}$ (Tl2201), this is established by angle-resolved photoemission (ARPES)~\cite{Peets2007} 
and angle-dependent magneto-resistance (ADMR)~\cite{Hussey2003} measurements, and quantum oscillations~\cite{Vignolle2008,Bangura2010}.
The measured cross-sectional area of the Fermi surface yields a carrier density
(per Cu atom)
$n = 1+p$.
The quantum oscillations also provide a measure of the carrier effective mass \mstar, whose value at $p = 0.29 \pm 0.02$ is 
\mstar~$= 5.2 \pm 0.4~m_0$~\cite{Bangura2010}.
Converting \mstar~to a specific heat coefficient $\gamma$ ($= C/T$ at $T \to 0$), via the relation 
$\gamma = 1.49$~(\mstar $/ m_0$) (in mJ/K$^2$mol),
yields 
$\gamma = 7.6 \pm 0.6$~mJ/K$^2$mol~\cite{Bangura2010}, in  agreement 
with the specific heat measured directly on a non-superconducting sample at $p = 0.33 \pm 0.02$, 
where
$\gamma = 6.5 \pm 1.0$~mJ/K$^2$mol~\cite{Wade1994}.
The data in Tl2201 are in excellent agreement with the two other cuprates whose specific heat has been measured at $p \simeq 0.3$,
namely La$_{2-x}$Sr$_x$CuO$_4$ (LSCO), where 
$\gamma = 6.9 \pm 0.7$~mJ/K$^2$mol at $p = 0.33$~\cite{Nakamae2003}
and
La$_{1.6-x}$Nd$_{0.4}$Sr$_x$CuO$_4$ (Nd-LSCO), where
$\gamma = 6.5 \pm 1.0$~mJ/K$^2$mol at $p = 0.36$~\cite{Michon2019}.
In summary, 
$\gamma \simeq 7$~mJ/K$^2$mol at $p \simeq 0.3$, a doping well above
\pstar~in all cases.
(Note that this value is 3 times larger than the value calculated from LDA band structure,
reflecting a significant mass enhancement due to electron correlations not captured by the calculations.)

\begin{figure}[t!]
\includegraphics[width=0.5\textwidth]{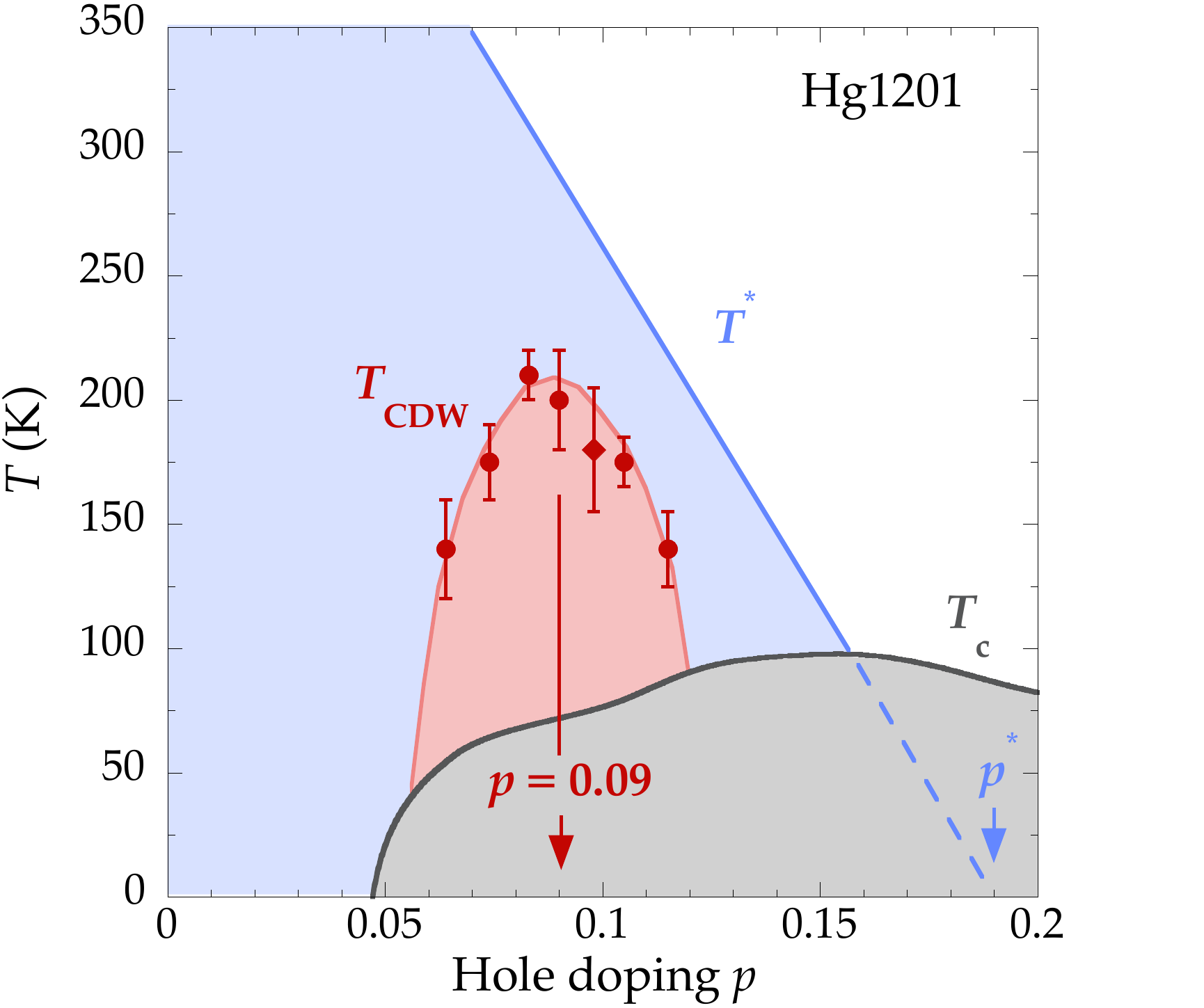}
\caption{
Temperature-doping phase diagram of Hg1201, showing the superconducting transition temperature 
\Tc~(grey line),
the pseudogap temperature \Tstar~(blue line), and the onset temperature for charge-density-wave (CDW) modulations
seen by resonant x-ray scattering (red circles) from Ref.~\onlinecite{Tabis2017} and references therein.
The red  vertical arrow indicates the doping of our sample 
($p \simeq 0.09$),
and the blue vertical arrow 
marks the pseudogap critical point $p^\star$, where \Tstar~extrapolates to zero.
%
}
\label{Fig1}
\end{figure}
%

The key question is what happens to that simple Fermi surface when doping is reduced below \pstar.
It is clearly transformed, but we still do not know exactly how.
ARPES studies on various cuprates show that states near $(\pi,0)$ are gapped~\cite{Matt2015}, leaving only 'Fermi arcs' at nodal locations in $k$-space,
also seen by scanning tunnelling spectroscopy (STM) in Ba$_{2}$Sr$_{2}$CaCu$_2$O$_8$ (Bi2212)~\cite{Fujita2014}.
Hall effect measurements on YBa$_2$Cu$_3$O$_y$ (YBCO)~\cite{Badoux2016}, Nd-LSCO~\cite{Collignon2017} 
and Tl2201~\cite{Putzke2019}
show a large drop
in the Hall number \nH, from \nH~$\simeq 1+p$ above \pstar~to \nH~$\simeq p$ below \pstar,
attributed to a drop in carrier density, also detected in thermal conductivity~\cite{Michon2018}.
It is tempting to interpret these various signatures in terms of a Fermi surface consisting
of four small closed nodal hole pockets containing a total of $p$ holes,
one side of which is detected as an arc in ARPES and STM.
But that remains to be demonstrated.

Associated with the Fermi surface transformation across \pstar~is a 10-fold drop in the DOS below \pstar,
first detected as a rapid reduction in the magnitude of the specific heat jump at \Tc~in YBCO~\cite{Loram1998}.
It appeared as though the opening of the pseudogap causes a loss of DOS.
Recently, a direct measurement of the normal-state specific heat at $T \to 0$ in Nd-LSCO and La$_{1.8-x}$Eu$_{0.2}$Sr$_x$CuO$_4$ (Eu-LSCO)
has suggested a different paradigm: before the DOS drops below \pstar, it first rises as $p \to$~\pstar~from above.
In other words, the DOS above and below \pstar~is more or less the same, but it goes through a large peak in between, at \pstar. 
Indeed in Nd-LSCO, where \pstar~$= 0.23$,
$\gamma \simeq 5$~mJ/K$^2$mol both at $p = 0.07$ and at $p = 0.40$, but
$\gamma \simeq 22$~mJ/K$^2$mol at $p = 0.24$~\cite{Michon2019}.
This peak displays the classic thermodynamic signature of quantum criticality, 
whereby $C/T \propto -$~log($1/T$) at \pstar~\cite{Michon2019}.
The Fermi surface transformation at \pstar~is therefore associated with a quantum critical point,
whose nature is as yet unknown.
Seen so far only in Nd-LSCO~\cite{Michon2019} and LSCO~\cite{Momono1994},
it is important to establish whether a peak in $C$ vs $p$ at \pstar, at $T \to 0$, is a generic property of cuprates.

In this Article, we explore this question with measurements of the specific heat in the cuprate material HgBa$_2$CuO$_{4 + \delta}$~(Hg1201),
at a doping $p \simeq 0.09$, well below \pstar~$\simeq 0.2$ (Fig.~1).
By applying a magnetic field of 35~T to suppress superconductivity, we access directly the normal-state $C(T)$ at low $T$,
and find a linear term $\gamma = 12 \pm 2$~mJ/K$^2$mol.
This is much larger than the value found in all other cuprates at $p \simeq 0.3 >$~\pstar.
If we require that the pseudogap in Hg1201 also causes a drop in DOS below \pstar, 
and assume that Hg1201 has the same $\gamma$ value as other cuprates at $p \simeq 0.3$,
then $C$ must peak at \pstar.
The very large $\gamma$ value we observe in Hg1201 also implies that the Fermi surface at $p \simeq 0.09$
includes more pieces than the one small pocket detected by quantum oscillations~\cite{Barisic2013,Chan2016}, forcing a revision of the current scenario
of Fermi-surface reconstruction by charge order~\cite{Chan2016,Harrison2011}.

%
\begin{figure}[t!]
\includegraphics[width=0.50\textwidth]{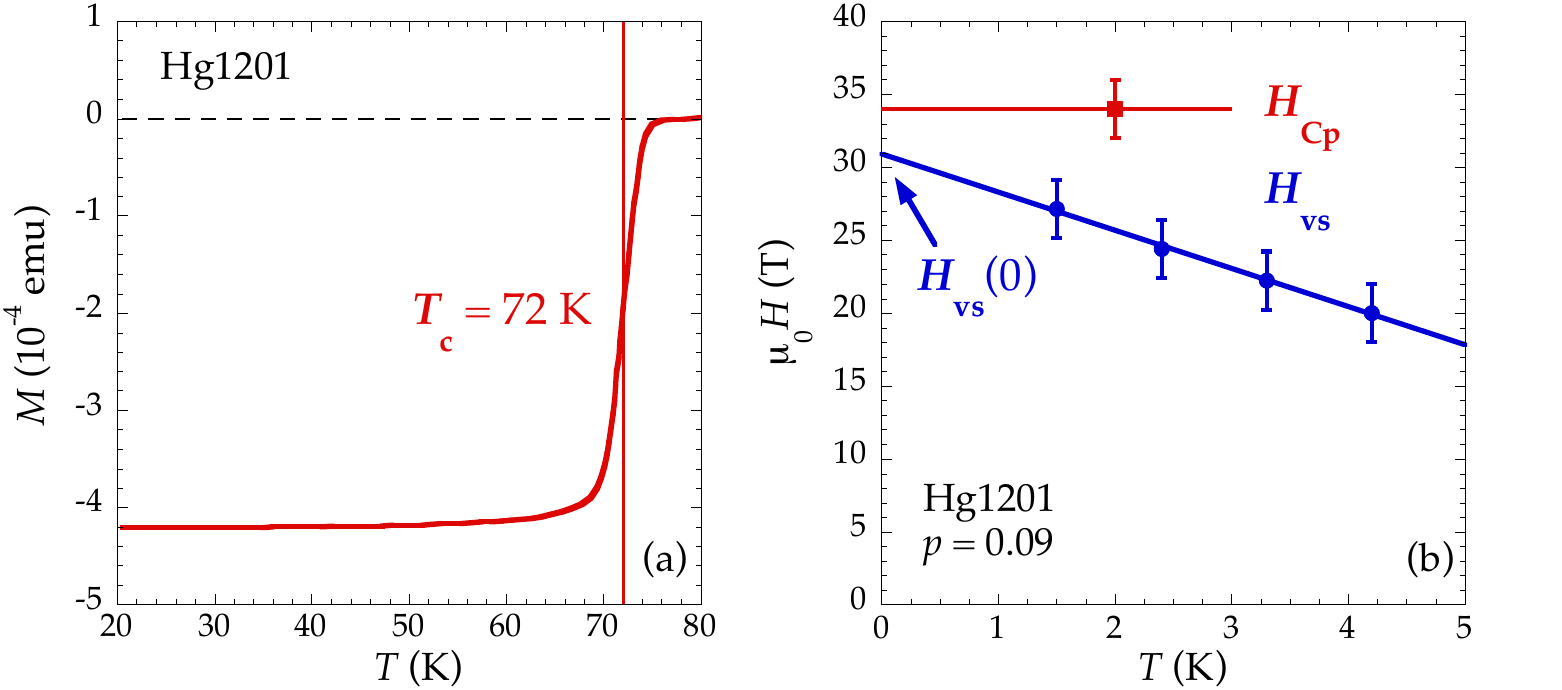}
\caption{
(a)
Zero-field cooled SQUID magnetization curve for our Hg1201 sample as a function of temperature, 
displaying a sharp superconducting transition at \Tc~$=72\pm2$~K, defined as the midpoint of the transition. 
(b)
Temperature dependence of the irreversibility field \Hvs$(T)$ (blue circles),
deduced from resistivity measurements
on a sample of Hg1201 with \Tc~$=72$~K
\cite{Barisic2013}. 
The vortex solid line extrapolates linearly to \Hvs~$= 31 \pm 2$~T (blue line), in agreement with
 the field above which the specific heat saturates at $T=2$~K, $H_{\rm Cp} = 34\pm 2$ T (red square and line). 
 This shows that our sample has a doping very similar to the doping at which quantum oscillations have been detected.}

\label{Fig2}
\end{figure}
%


\section{II.~Methods}

Our single crystal of Hg1201 was grown using a self-flux technique~\cite{Legros2019}. Its mass is $m \simeq 1.1$~mg.
 It was annealed in a vacuum of $3 \times 10^{-1}$~mbar at 275~$^\circ$C during 67 hours,
 to produce a superconducting transition temperature \Tc~=~72~K,
defined as the mid-point of the drop in magnetization measured in a SQUID magnetometer, with a field of 10~Oe 
(see Fig.~\ref{Fig2}a).
The estimated hole concentration (doping) for such a \Tc~value is $p \simeq 0.09$~(Fig.~1).

%
\begin{figure}[t!]
\includegraphics[width=0.50\textwidth]{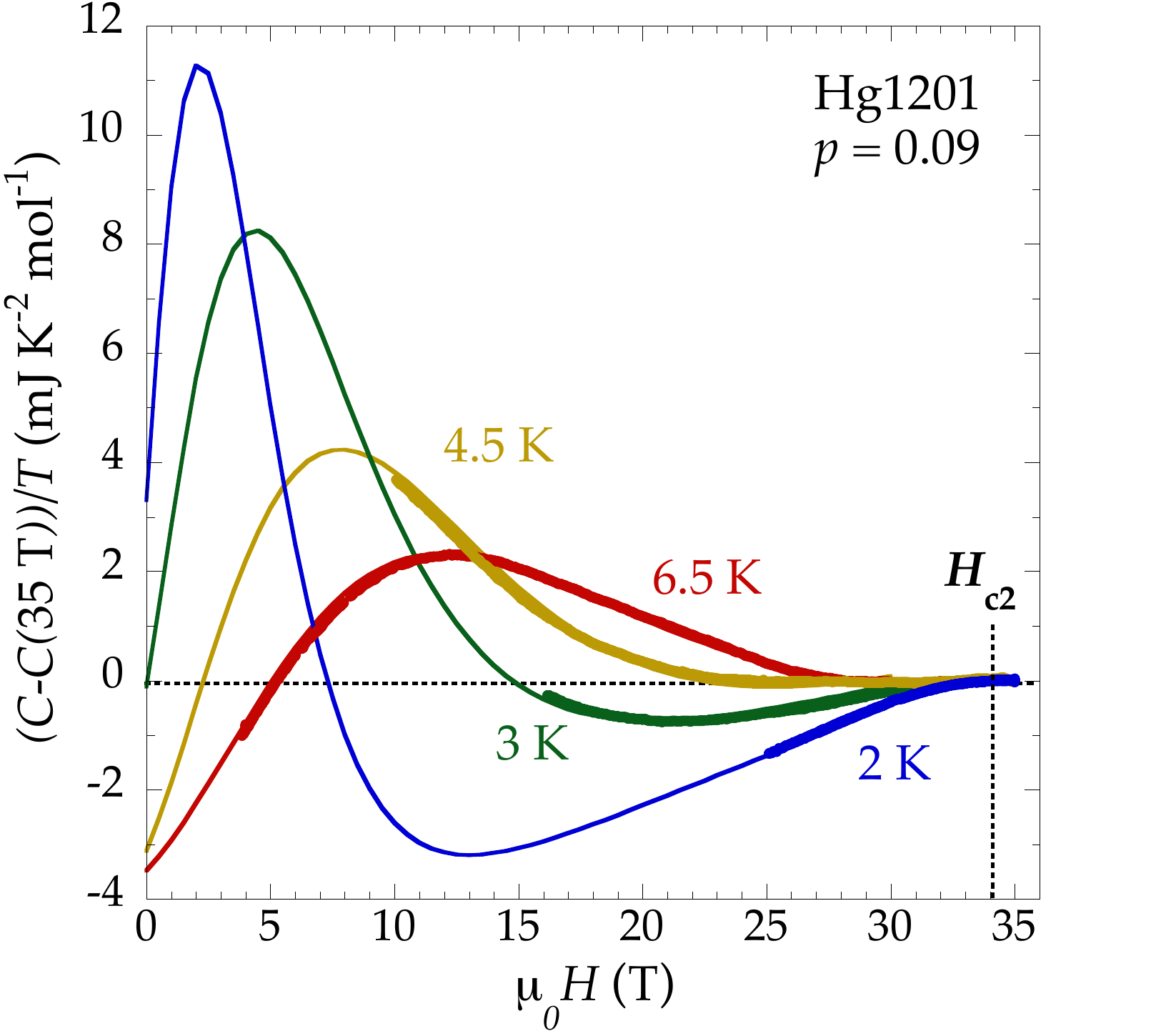}
\caption{
Specific heat $C$ of our Hg1201 sample as a function of magnetic field $H$, plotted as $C/T$ vs $H$, for four temperatures as indicated.
A constant term is subtracted from each isotherm, namely the value of $C/T$ at $H = 35$~T, 
which is plotted in Fig.~4.
The vertical dashed line marks the upper critical field $ H_{\rm C_p}=34 \pm 2$~T, defined as the field above which $C$ vs $H$ has saturated.
The thin lines indicate for each temperature the field dependence expected for a standard two level Schottky contribution with a gap varying as $\Delta/k_{\rm B}= 2.5+1.2\times H$, 
on top of a linear field dependence of the electronic specific heat in the superconducting state (up to 25~T).
}
\label{Fig3}
\end{figure}
%

The specific heat was measured using an AC micro-calorimetry technique described in Ref.~\onlinecite{Michon2019}.  
The total heat capacity, $C$, was obtained through the equation $C = P_{\rm ac} \sin(-\phi) / 2\omega |T_{\rm ac}|$, 
where $P_{\rm ac}$ is a periodically modulated heating power,  $\phi$ is the thermal phase shift and $T_{\rm  ac}$ the induced temperature oscillations. 
A miniature Cernox resistive chip was split into two parts and attached to a small copper ring with PtW(7\%) wires. 
The first half was used as the heater delivering $P_{\rm ac}$ and the second half was used to record the temperature $T_{\rm  ac}$.
In order to subtract the heat capacity of the sample mount (chip + a few $\mu$g of Apiezon grease used to glue the sample onto the back of the chip), the empty chip (with grease) was measured prior to the sample measurements. 
A precise \textit{in situ} calibration and corrections of the thermometers in magnetic field were included in the data treatment. 
This technique enabled us to obtain the absolute value of the specific heat of miniature single crystals with an accuracy better than $\sim 5~\%$~below 10~K, 
as checked from measurements on ultrapure copper \cite{Michon2019}.  

A magnetic field was applied normal to the CuO$_2$ planes (along the $c$ axis) to suppress superconductivity.
The zero temperature upper critical field \Hc~can be obtained from resistance measurements of the vortex-solid field \Hvs$(T)$, 
defined as the field below which the sample resistance is zero \cite{Grissonnanche2014}.
By extrapolating \Hvs$(T)$ to $T \to 0$, we get \Hc~=~\Hvs (0). From the data of Ref.~\onlinecite{Barisic2013}, on a sample of Hg1201 with 
a very similar \Tc~$(=72$~K), 
we find \Hc~$= 31 \pm 2$~T
(Fig.~\ref{Fig2}b).
We shall see that this is nicely consistent with the \Hc~value obtained from the saturation of $C$ vs $H$ in our own sample 
{\it i.e.} $H_{\rm c2}\approx H_{\rm C_p} =34\pm2$~T (Fig.~\ref{Fig2}b).
Given that \Hc~varies rapidly with $p$ near $p = 0.1$ (in YBCO and presumably also in Hg1201), this matching of \Hc~and \Tc~values 
confirms that our specific heat data and the quantum oscillation data of Ref.~\onlinecite{Barisic2013} 
are being compared at the same doping.


\begin{figure}[t!]
\includegraphics[width=0.45\textwidth]{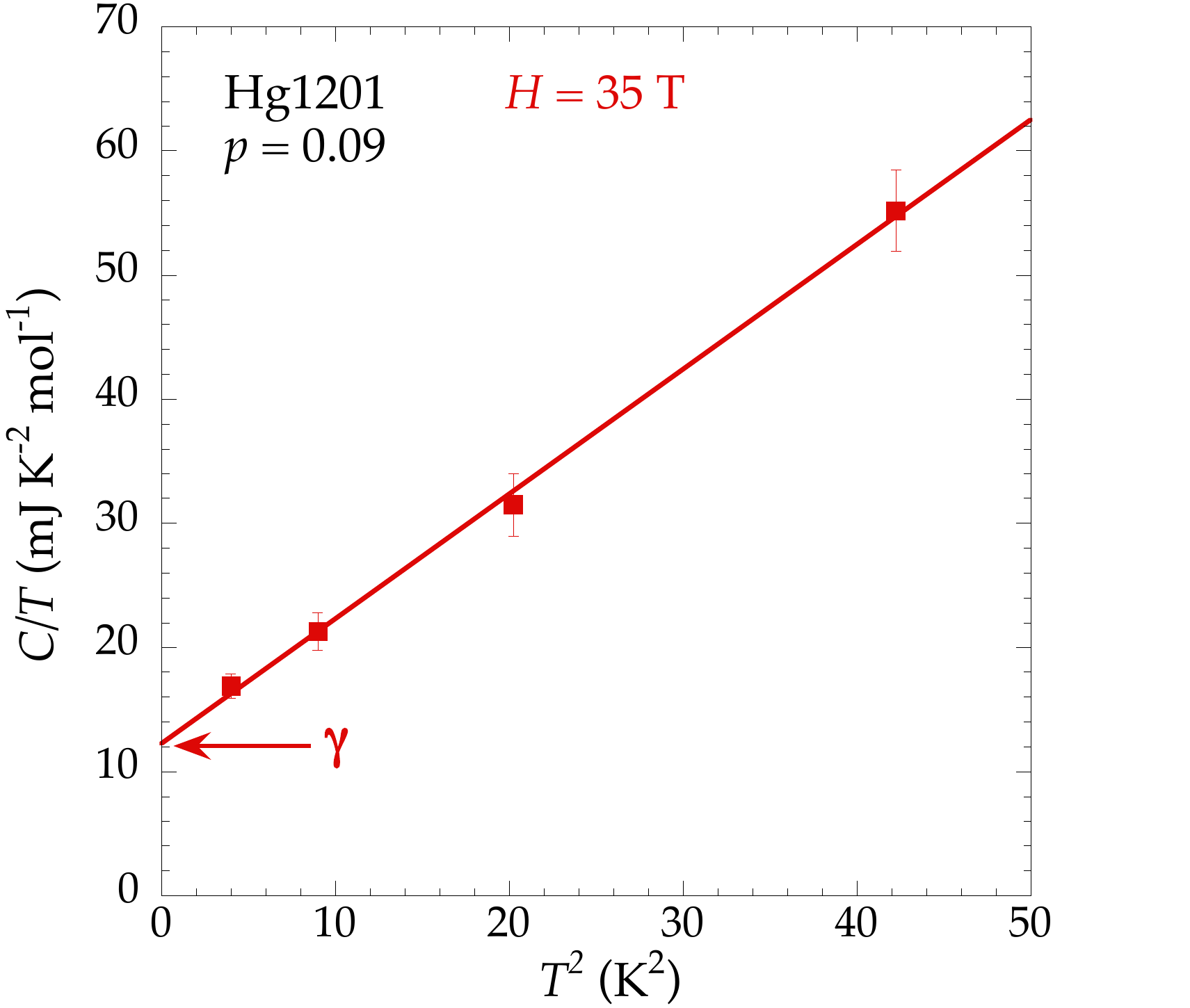}
\caption{
Normal-state specific heat at $H = 35$~T, for the four isotherms of Fig. \ref{Fig3}, plotted as $C/T$ vs $T^2$ (red squares).
The solid red line is a linear fit of the four data points to 
$C/T = \gamma + \beta T^2$, 
giving $\gamma = 12 \pm 2$~mJ/K$^2$mol and $\beta = 1.0 \pm 0.1$~mJ/K$^4$mol. 
}
\label{Fig4}
\end{figure}
%



\section{III.~Results}

In Fig.~3, we show the specific heat $C$ of our Hg1201 sample, plotted as $C/T$ vs $H$, at 4 different temperatures:
$T = 2.0, 3.0, 4.5,$~and 6.5~K.
For clarity, the data have been shifted to zero at 35~T by subtracting the value of $C/T$ at $H = 35$~T,
which is itself plotted in Fig.~4.
In Fig.~3, a Schottky anomaly is clearly visible below $\sim 30$~T for the highest temperature (6.5~K). \
This Schottky contribution can be well described by
the standard expression
$C_{\rm Schottky} \propto (\Delta/k_{\rm B}T)^2\exp(\Delta/k_{\rm B}T)/(1+\exp(\Delta/k_{\rm B}T))^2$ with $\Delta/k_{\rm B} \approx 2.5+1.2\times H$ (K) (thin solid lines in Fig.~3), 
in agreement with the gap value previously inferred by Kemper \cite{Kemper2014}. 
As expected, this Schottky contribution
moves progressively to lower fields with decreasing temperature. At $T = 4.5$~K, it is negligible above 25~T 
and at our base temperature of 2~K, the data are free of Schottky contribution above 25~T.
At $T=2$~K, 
the increase in $C/T$ vs $H$ 
reflects the suppression of superconductivity,
which is complete by 35~T;
$C/T$ vs $H$ has reached saturation for fields above $H_{\rm Cp}=34 \pm 2$~T.
The value of $C/T$ at $H = 35$~T is therefore the normal-state value, 
free of any Schottky contribution,
plotted as $C/T$ vs $T^2$ in Fig.~4.

%
\begin{figure}[t!]
\includegraphics[width=0.47\textwidth]{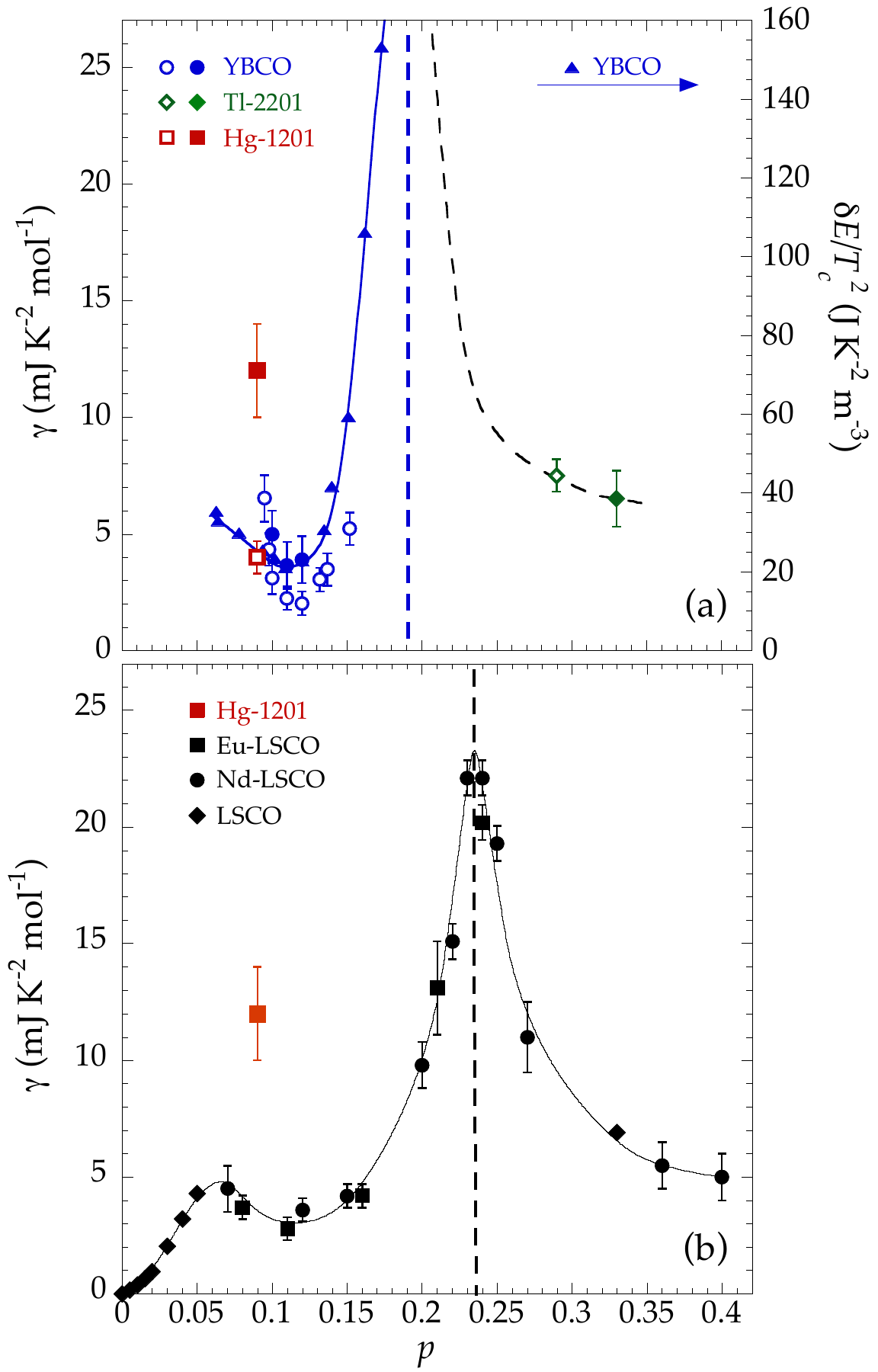}
\caption{
(a) Normal-state specific heat coefficient $\gamma$ (= $C/T$ at $T \to 0$) in three hole-doped cuprates:
Hg1201 at $p \simeq 0.09$ (red square, left axis, 
from Fig.~4);
YBCO at $p = 0.10$, 0.11 and 0.12 (solid blue circles, left axis~\cite{Kacmarcik2018});
non-superconducting Tl2201 at $p = 0.33 \pm 0.02$ (solid green diamond, left axis~\cite{Wade1994}).
Also shown are the values of $\gamma$ obtained from the effective mass \mstar~measured by quantum oscillations (Eq.~1) in 
Hg1201 at $p \simeq 0.09$ (open red square, left axis~\cite{Chan2016}); YBCO at $0.08 < p < 0.16$ (open blue circles, left axis~\cite{Sebastian2010,Ramshaw2015}); 
Tl2201 at $p = 0.29 \pm 0.02$ (open green diamond, left axis~\cite{Bangura2010}).
The units for $C$ are expressed per mole of planar Cu. The blue solid line is a guide to the eye. 
The blue dashed line marks the pseudogap critical point \pstar~in YBCO. The black dashed line is a guide to the eye. 
(b) Normal-state specific heat coefficient $\gamma$ in four hole-doped cuprates:
Hg1201 at $p \simeq 0.09$ (red square, from Fig.~4);
LSCO (black diamonds) at $p < 0.06$~\cite{Komiya2009} and at $p = 0.33$~\cite{Nakamae2003};
Nd-LSCO (black circles~\cite{Michon2019}); Eu-LSCO (black squares~\cite{Michon2019}).
The data points at $p = 0.20$, 0.21, 0.22, 0.23, 0.24 and 0.25 are not $\gamma$ but rather 
$C_{\rm el}/T$ at $T = 0.5$~K, where $C_{\rm el}$ is the normal-state electronic specific heat~\cite{Michon2019}.
The black dashed line marks \pstar~in Nd-LSCO \cite{Michon2019,Collignon2017}. 
}
\label{Fig5}
\end{figure}
%


The 35~T normal state data are well described by a linear fit, $C/T = \gamma + \beta T^2$, 
with $\gamma = \gamma_{\rm N}= 12 \pm 2$ mJ/K$^2$mol and $\beta = 1.0 \pm 0.1$~mJ/K$^4$mol,
with error bars that combine the uncertainty on the absolute value of $C$ and the uncertainty on the fit. 
Our values for $\gamma$ and $\beta$ are in excellent agreement with those previously obtained by Kemper
on an underdoped crystal of Hg1201 with \Tc~$=72$~K~\cite{Kemper2014}. 
Note that in a $d$-wave superconductor, a non-zero 
residual Sommerfeld coefficient $\gamma_{\rm R}$ is usually observed due to disorder-induced pair-breaking effects. 
The value of $\gamma_{\rm R}$  then strongly depends on disorder, and slight variations in the level of disorder from sample to sample will result in large variations in $\gamma_R$, as reported previously \cite{Kemper2014}. 
However, disorder does not affect the normal state $\gamma_{\rm N}$ , and our observation of 
the same large $\gamma_{\rm N}$  value in two separate studies
confirms that it is an intrinsic electronic property of Hg1201.


\section{IV.~Discussion}

%
In Fig.~5, we compare our value of $\gamma$ in Hg1201 at $p = 0.09$ to $\gamma$ values previously measured in other hole-doped cuprates
(in the normal state without superconductivity).
The values reported so far at dopings close to $p = 0.1$ are listed in Table~I (where units are per mole of planar Cu).
We see that $\gamma$ in Hg1201 (12~mJ/K$^2$mol) is significantly larger than in 
YBCO (5~mJ/K$^2$mol)~\cite{Marcenat2015,Kacmarcik2018}, 
LSCO (5~mJ/K$^2$mol)~\cite{Momono1994},  Nd-LSCO (4~mJ/K$^2$mol)~\cite{Michon2019}, and Eu-LSCO (3~mJ/K$^2$mol)~\cite{Michon2019}.


\begin{table}[t]
\centering
 \addtolength{\leftskip} {-2cm}
 \addtolength{\rightskip}{-2cm}

\begin{tabular}{|c|c|c|c|c|c|}

\hline
Material & Doping & State  &  $\gamma$ & Ref.  \\

 &  &   &  (mJ/K$^2$mol )&   \\

\hline


 Hg1201 & 0.09  & SR-CDW &  $12 \pm 2$ &   this work   \\


 YBCO & 0.10  & LR-CDW &  $5 \pm 1$ & \cite{Kacmarcik2018}   \\

 LSCO & 0.10  & CDW + SDW  &  $5 \pm 1$ & \cite{Momono1994}   \\

 Eu-LSCO & 0.11 & CDW + SDW  &  $2.8 \pm 0.5$ & \cite{Michon2019}   \\

 Nd-LSCO & 0.12  & CDW + SDW  &  $3.6 \pm 0.5$ & \cite{Michon2019}   \\

\hline

 LSCO & 0.33  & FL  &  $6.9 \pm 0.7$ & \cite{Nakamae2003}   \\

 Tl2201 & 0.33  & FL  &  $6.5 \pm 1.0$ & \cite{Wade1994}   \\

 Nd-LSCO & 0.36  & FL  &  $6.5 \pm 1.0$ & \cite{Michon2019}   \\

\hline
\end{tabular}

\caption{
Residual linear term $\gamma$ in the specific heat of various hole-doped cuprates,
measured in the normal state as $C/T$ in the limit $T \to 0$, both in the underdoped regime at $p \simeq 0.1$ (top group)
and in the strongly overdoped regime at $p > 0.3$ (bottom group).
The units for $\gamma$ are expressed per Cu atom in the CuO$_2$ planes.
In the absence of superconductivity, the ground state is either
long-range (LR) 3D CDW order (in YBCO), short-range (SR) CDW correlations (in Hg1201),
combined CDW and SDW modulations (stripe order), or a Fermi liquid (at $p >\,0.3$).
}

\label{T1}
\end{table}


\subsection{A.~Fermi surface and CDW order}

At $p \simeq 0.1$, various factors will affect the DOS in the normal state. 
First, the pseudogap reduces the DOS (see discussion below).
Secondly, in all the former materials there is some form of charge-density-wave (CDW) order (or correlations) at $p \simeq 0.1$,
detected by x-ray diffraction in
Hg1201~\cite{Tabis2017},
YBCO~\cite{Blanco-Canosa2014,Hucker2014},
LSCO~\cite{Croft2014},
Nd-LSCO~\cite{Niemoller1999}, and
Eu-LSCO~\cite{Fink2011},
amongst others~\cite{Keimer2015}.
This CDW order causes a reconstruction of the Fermi surface,
detected as a change of sign in the Hall and Seebeck coefficients, from positive at high temperature to negative 
at low temperature, in
Hg1201~\cite{Doiron-Leyraud2013},
YBCO~\cite{LeBoeuf2007,Chang2010,LeBoeuf2011}, 
LSCO~\cite{BadouxPRX2016}, 
Nd-LSCO~\cite{Nakamura1992}, and 
Eu-LSCO~\cite{Laliberte2011}.
This reconstruction reduces the DOS further, beyond the effect of the pseudogap, as indicated by the fact 
that $\gamma$ vs $p$ has a local minimum at $p \simeq 0.12$ in 
YBCO~\cite{Kacmarcik2018}, 
LSCO~\cite{Momono1994}, 
Nd-LSCO~\cite{Michon2019}, and 
Eu-LSCO~\cite{Michon2019},
where CDW order is strongest.
There is also a local minimum in the upper critical field, \Hc~vs $p$, and in the associated condensation energy~\cite{Grissonnanche2014}.

In YBCO, at the high fields used to access the normal-state $\gamma$ ($H > 25$~T), there is a long-range,
unidirectional 3D CDW order~\cite{Gerber2015,Chang2016}, not observed so far in any other cuprate.
In particular, this long-range 3D order has not been seen in Hg1201. Although the comparative impact on the Fermi surface and associated DOS of this 3D order $vs$ the short-range 2D order is not yet clear, it is conceivable that the smaller $\gamma$ value in YBCO (5~mJ/K$^2$mol) $vs$ Hg1201 (12~mJ/K$^2$mol) has to do with the difference in their CDW ordering. (Note also that $\gamma$ in YBCO rises fast below $p = 0.11$~\cite{Kacmarcik2018}, and so will become larger than 5~mJ/K$^2$mol at $p < 0.10$.)

A third factor that should affect the DOS of the normal state is spin order,
which occurs in addition to charge order in LSCO, Nd-LSCO and Eu-LSCO, at $p \simeq 0.1$~({\it e.g.}~\cite{Chang2008}),
but {\it not} in Hg1201 or YBCO~\cite{Wu2011}.
Indeed, having spin order in addition to charge order is expected to modify the way in which the Fermi surface is reconstructed~\cite{Millis2007}.
It is therefore conceivable that the larger $\gamma$ value in 
Hg1201 (12~mJ/K$^2$mol) $vs$
LSCO (5~mJ/K$^2$mol), 
Nd-LSCO (4~mJ/K$^2$mol) or
Eu-LSCO (3~mJ/K$^2$mol), 
could be due to the presence of spin ordering in the latter materials.

A fourth factor 
that could affect the DOS 
is the proximity of a van Hove singularity. 
Such a singularity is present in 
hole-doped 
cuprates as the Fermi surface goes from hole-like to electron-like upon doping. 
However, ARPES data on Hg1201 \cite{vishik2014} show that the van Hove singularity in Hg1201 is located at a doping well above optimal doping,
 in accordance with a tight binding model that predicts $p_{\rm vHs} \gg 0.19$. 
(In Nd-LSCO, $p_{\rm vHs} \approx 0.23$ and 
yet a
$ \gamma$ value as low as 4~mJ/K$^2$~mol is reported at $p=0.12$ \cite{Michon2019}.)
Therefore,
given that $p_{\rm vHs} \gg 0.09$ in Hg1201,
we expect that the $\gamma$ value measured at $p = 0.09$ in Hg1201 is only slightly affected by the van Hove singularity,
and the large value of 12~mJ/K$^2$~mol is certainly not the result of this singularity. 

Our current knowledge of the Fermi surface of YBCO and Hg1201 comes mostly from quantum oscillations,
detected in YBCO at dopings from $p \simeq 0.09$ to  $p \simeq 0.15$~\cite{Doiron-Leyraud2007,Sebastian2010,Ramshaw2015}
and in Hg1201 at $p \simeq 0.09$~\cite{Barisic2013,Chan2016}.
These oscillations provide a direct way to measure the effective mass \mstar, for each (closed) piece of the Fermi surface.
In 2D, a sum rule requires that the various values of \mstar~must add up to the specific heat $\gamma$~\cite{Mackenzie1996}:
\begin{equation}
\gamma = {\rm prefactor} \times \sum n_\mathrm{i}~(m_\mathrm{i}/m_0)~~,
\end{equation}
where $n_i$ is the number of equivalent pockets in the Brillouin zone,
$m_i$ is the mass \mstar~of each independent pocket, $m_0$ is the electron mass,
and the prefactor is equal to 1.47 for YBCO and 1.49 for Hg1201, when $\gamma$ is expressed in 
mJ/K$^2$mol (per mole of planar Cu).

In YBCO, at least 4 different frequencies have been resolved~\cite{Doiron-Leyraud2015},
the interpretation of which is still debated.
At $p = 0.10$, the dominant frequency $F = 530$~T has a mass \mstar~$= 1.9 \pm 0.1~m_0$~\cite{Doiron-Leyraud2007,Sebastian2010,Ramshaw2015},
which yields $\gamma = 1.47 \times 1.9 = 2.8 \pm 0.2$~mJ/K$^2$mol (assuming one pocket per CuO$_2$ plane).
This is significantly smaller than the measured  $\gamma = 5 \pm 1$~mJ/K$^2$mol~(Fig.~3a).
Possible explanations for the missing mass include: the presence of a second closed pocket (with $n_2 m_2 = 1.5~m_0$)~\cite{Doiron-Leyraud2015};
the open band associated with the CuO chains of YBCO; an open piece of the Fermi surface associated with the CuO$_2$ planes.

In Hg1201, the situation is much simpler: there is only one CuO$_2$ plane per unit cell, no chains and 
only a single frequency is observed, giving $F = 850$~T and  \mstar~$= 2.7 \pm 0.1~m_0$, for a sample with \Tc~$=71$~K~\cite{Chan2016}.
If this single frequency corresponds to one pocket per CuO$_2$ plane ($n_1 = 1$),
then $\gamma = 1.49 \times 2.7 = 4.0 \pm 0.2$~mJ/K$^2$mol.
Similarly, in Ref.~\onlinecite{Barisic2013},
$F = 840$~T and \mstar~$= 2.45 \pm 0.15~m_0$, for a sample with \Tc~$=72$~K, 
giving $\gamma = 3.7 \pm 0.2$~mJ/K$^2$mol. 
(As discussed above in relation to Fig.~2b, with the same values of \Tc~and \Hc,
our sample and the sample of ref.~\onlinecite{Barisic2013} have the very same doping).
%
%
Having a value 3 times smaller than the measured $\gamma$ then immediately implies
that the Fermi surface of Hg1201 includes pieces beyond the small closed pocket that gives rise to the quantum oscillations.
Therefore, the main scenario proposed so far for Hg1201~\cite{Chan2016}, whereby the Fermi surface consists of a single electron-like pocket 
per CuO$_2$ plane resulting from a reconstruction by biaxial CDW order~\cite{Harrison2011}, is ruled out.

The additional pieces of Fermi surface can either be other closed pockets, as yet undetected in quantum oscillation measurements,
or open bands, undetectable in such measurements. If closed pockets, their total mass must be twice that of the measured mass (\mstar~$= 2.7~m_0$).
By comparison, in YBCO the total 'missing mass' is only 75\% of the main mass (\mstar~$= 1.9~m_0$).
If open bands, these must represent a significant fraction of the total DOS.
Note that open bands have been proposed in the context of a reconstruction by uniaxial CDW order~\cite{Yao2011}.
Either way, a major rethinking of the Fermi surface of Hg1201 is necessary, and more generally of all underdoped cuprates.


\subsection{B.~Pseudogap and peak in $\gamma$ vs $p$}

Irrespective of what is the correct Fermi surface for Hg1201 at $p \simeq 0.1$, the striking fact remains
that its measured $\gamma$ (12~mJ/K$^2$mol) is significantly larger than what is measured in 
the overdoped regime at $p > 0.3$, in various single-layer cuprates (Table~I):
Tl2201 (6.5~mJ/K$^2$mol)~\cite{Wade1994};
LSCO (6.9~mJ/K$^2$mol)~\cite{Nakamae2003};
Nd-LSCO (6.5~mJ/K$^2$mol)~\cite{Michon2019}.
How is that possible if the opening of the pseudogap below \pstar~$\simeq 0.2$ (Fig.~1) causes a loss of DOS ? 
A first explanation could be that this particular singe-layer cuprate has a $\gamma$ value for $p >$~\pstar~much higher than 
the $\gamma$~value measured
in the other three single-layer cuprates at $p = 0.3$, i.e. LSCO, Nd-LSCO and Tl2201. 
However, we
cannot think of any physical reason for that, in a regime where properties obey Fermi liquid theory and the Fermi surface is properly given by band structure calculations.
A second, more natural
explanation is that $\gamma$ rises in going from $p \simeq 0.3$ to $p =$~\pstar,
and then drops in going from $p =$~\pstar~to $p \simeq 0.09$, upon entering the pseudogap phase.
In other words, $\gamma$ peaks at \pstar.
Such a peak has been measured directly in both LSCO (with superconductivity removed by introduction of Zn impurities)~\cite{Momono1994}
and in Nd-LSCO (with superconductivity removed by application of a magnetic field)~\cite{Michon2019}.

In Nd-LSCO, the electronic specific heat $C_{\rm el}$ peaks sharply at \pstar~=~0.23 (Fig.~3b).
At $p = 0.24$, $C_{\rm el}/T$ increases logarithmically as $T \to 0$, to reach $C_{\rm el}/T \simeq 21$~mJ/K$^2$mol
at $T = 0.5$~K~\cite{Michon2019}.
In YBCO, there are no direct measurements of the normal-state $\gamma$ above $p = 0.12$,
because the magnetic fields needed to suppress superconductivity when $p >\,0.12$ rapidly exceed 45~T~\cite{Grissonnanche2014}.
It is nevertheless clear~\cite{Proust2019}, from indirect measurements~\cite{Loram1998,Grissonnanche2014}, 
that the DOS increases dramatically in going from $p = 0.12$ to $p =$~\pstar~=~0.19.
For example, in standard BCS theory, $\gamma \propto \delta E$/\Tc$^2$,
where $\delta E$ is the condensation energy~\cite{Proust2019}.
An estimate of $\delta E$ in YBCO
via measurements of the upper (\Hc) and lower ($H_{\rm c1}$) critical fields,
finds that $\delta E$/\Tc$^2$ increases by a factor 6.5 in going from $p = 0.10$ to $p = 0.18$~\cite{Grissonnanche2014} (Fig.~3a).
Given that $\gamma = 5$~mJ/K$^2$mol at $p = 0.10$~(Table~I), this implies that
$\gamma \simeq 35$~mJ/K$^2$mol at $p =$~\pstar~(Fig.~3a) -- 
a value 3 times larger than $\gamma$ in Hg1201 at $p \simeq 0.09$ and 5 times larger than $\gamma$ at $p \simeq 0.33$.
We propose that if Hg1201 could be doped up to $p \simeq 0.3$ and its normal-state $\gamma$ could be measured
across \pstar, one would find that $\gamma \simeq 7$~mJ/K$^2$mol at $p = 0.3$
and $\gamma \simeq 30$~mJ/K$^2$mol at $p =$~\pstar.


\section{V.~Conclusion}

By applying a magnetic field of 35~T to the single-layer cuprate Hg1201 at a doping $p \simeq 0.09$,
we have suppressed its superconductivity and measured its normal-state specific heat $C$.
Extrapolating $C/T$ to $T = 0$ yields $\gamma = 12 \pm 2$~mJ/K$^2$mol.
%
%
This high value of $\gamma$ has two major implications.
First, it is significantly larger than the value measured in overdoped cuprates outside the pseudogap phase,
where $\gamma \simeq 7$~mJ/K$^2$mol.
Given that the pseudogap causes a loss of density of states, this implies that $\gamma$ must  
peak between $p \simeq 0.1$ and $p \simeq 0.3$, namely at (or near) the  critical doping \pstar~where
the pseudogap phase is expected to end (\pstar~$\simeq 0.2$) -- 
as indeed found in LSCO and Nd-LSCO.
%
%
Secondly, the high $\gamma$ value implies that the Fermi surface of Hg1201 must consist of more than the single electron-like 
pocket detected by quantum oscillations in Hg1201 at $p \simeq 0.09$,
whose effective mass
yields only $\gamma = 4.0 \pm 0.2$~mJ/K$^2$mol.
This missing mass imposes a revision of the current scenario for how pseudogap and charge order 
respectively transform and reconstruct the Fermi surface of cuprates.



\section{Acknowledgments}


L.T. acknowledges support from the Canadian Institute for Advanced Research (CIFAR) as a CIFAR Fellow
and funding from the Natural Sciences and Engineering Research Council of Canada (NSERC; PIN:123817), 
the Fonds de recherche du Qu\'ebec -- Nature et Technologies (FRQNT), 
the Canada Foundation for Innovation (CFI), 
and a Canada Research Chair. 
This research was undertaken thanks in part to funding from the Canada First Research Excellence Fund. 
Part of this work was funded by the Gordon and Betty Moore Foundation's EPiQS Initiative (Grant GBMF5306 to L.T.).


%

\end{document}